\documentclass[
showpacs,preprintnumbers,amsmath,amssymb]{revtex4}
\usepackage{amsmath}
\usepackage{graphicx}
\begin{document}

\title{INFLATIONARY UNIVERSE WITH A VISCOUS  FLUID AVOIDING SELF-REPRODUCTION}

\author{I. Brevik$^{1}$\footnote{E-mail:iver.h.brevik@ntnu.no} }

\medskip

\affiliation{Department of Energy and Process Engineering, Norwegian University
of Science and Technology, N-7491 Trondheim, Norway}

\author{E. Elizalde$^{2}$\footnote{E-mail: elizalde@ieec.uab.es}}

\medskip

\affiliation{Consejo Superior de Investigaciones Cientificas, ICE/CSIC-IEEC, Campus UAB, Carrer de Can Magrans, s/n, 08193 Bellaterra (Barcelona), Spain}

\author{ V. V. Obukhov}

\medskip

\affiliation{Tomsk State Pedagogical University, Kievskaja Street 60, 634050 Tomsk, Russia}

\author{A. V. Timoshkin$^{{3},{4}}$\footnote{E-mail:alex.timosh@rambler.ru}}

\medskip

\affiliation{$^{3}$Tomsk State Pedagogical University, Kievskaja Street 60, 634050 Tomsk, Russia}
\affiliation{$^{4}$Tomsk State University of Control Systems and Radioelectronics, Lenin Avenue 40, 634050 Tomsk, Russia}

\begin{abstract}
We consider a universe with a bulk viscous cosmic fluid, in a flat Friedmann-Lemaitre-Robertson-Walker geometry. We derive the conditions for the existence of inflation, and those which at the same time prevent the occurrence of self-reproduction. Our theoretical model gives results which are in perfect agreement with the most recent data from  the PLANCK surveyor.
\end{abstract}

 \maketitle

 \today

\section{Introduction}

Recent observational data from  the Planck satellite concerning cosmic inflation \cite{ade13,ade13a} have enabled us to make a more detailed picture of the inflationary period of the universe evolution. The inflationary theory describes the very early and intermediate stage of this evolution. The inflationary stage is known to be extremely short, but the universe expansion becomes  exponentially large during this epoch.

However, the usual inflationary theories have inherent problems, such as multiversity, predictability, and initial condition issues. Self-reproduction of the universe, meaning that the inflationary process has no way to finish, can be also seen as a major problem; the inflation  would never end. Recently, an inflationary scenario which avoids the self-reproduction problem has  been proposed in \cite{mukhanov14}.

There exist several different regimes which are possible in inflationary cosmology. We will here consider what we believe is the simplest possibility for inflation without self-reproduction. In such scenario the universe is not stationary.

As shown in \cite{nojiri15},  $F(R)$ inflation without self-reproduction may be formulated analogously to the corresponding scalar models considered in \cite{mukhanov14} (for a review of inflation in $F(R)$ gravity, see \cite{bamba:2015uma}).  A description of an inflationary universe assuming a perfect fluid model and $F(R)$ gravity was given in \cite{bamba14}, including comparison with observational data. A general review of inflationary cosmology and associated problems involving multiversity and initial conditions was given in \cite{linde}, while additional relevant literature on the subject we discuss here can be found in \cite{addrefs1}.

In order to avoid  self-reproduction in the inflationary universe the thermodynamic parameter $\omega(N)$ in the  equation of state (EoS) is bound to satisfy the following requirements:
\begin{align}
& a)~ 1+\omega(N)=1 ~{\rm at} ~N=1~{\rm (to~ have~ graceful~ exit)},\notag \\
& b)~ 1+\omega(N) \leq 2/3~{\rm at} ~N=N_m~{\rm (to ~solve~ the~ initial~ condition~ problem)}, \notag \\
& c) ~1+\omega(N) \ll 1~{\rm for}~1<N<N_m~{\rm (period~ of~ inflation)}, \notag \\
& d)~ 1+\omega(N) > \epsilon(N)~{\rm for}~1<N<N_m~{\rm (no~ selfreproduction)}. \label{1}
\end{align}
Note that all these given conditions for avoiding self-reproduction are referred to Einstein's gravity background (see Ref. [4]). The origin of the requirements (1) are carefully explained in Ref. [4]. Concerning the first of them, one should observe that, from the very beginning of inflation a small (but nonvanishing) deviation of the EoS from the cosmological constant is compulsory (first of Eqs. (1)), because otherwise inflation would never end. To describe the change of this deviation with time, the number of e-folds, $N$, left up to the end of inflation is used, which plays the role of time.  At the beginning of inflation one has $N=N_m$, while at its end, $N \sim 1$. The scale factor $a$ varies as $a=a_0e^{-N}$, with $a_0$ a constant. One expects $N$ to be moderately large, namely $N<70$.  Moreover, $\epsilon(N)$ is the non-dimensional energy density in Planck units, related to the energy density $\rho$ in common units via $\epsilon(N) \sim k^4 \rho(N)$, with $k^2=8\pi G$. We will not repeat here the discussion leading to the rest of the conditions (1), which can be found in Ref. [4].

Viscous cosmology is a subject that has attracted considerable attention recently. The inclusion of viscosity means physically that one works up to first order deviations from thermal equilibrium. For instance, the inclusion of bulk viscosity (the influence of shear viscosity being absent because of spatial isotropy) may be of importance for the occurrence of the big rip future singularity \cite{brevik06,nojiri05}, as well as for singularities of the so-called  II, III, and IV types \cite{nojiri05,brevik10}, and for the  dark energy/dark matter coupling \cite{brevik14}.

It should be noted that viscous fluids can be considered as a particular class of generalized fluids (fluids with inhomogeneous or time dependent  EoS, see \cite{myrzakulov15,nojiri05,capozziello06,nojiri06,nojiri07}). Inhomogeneous fluid cosmology may also be interpreted as a kind of modified gravity \cite{nojiri07a,nojiri11}. Various examples of inhomogeneous viscous fluids were investigated in \cite{bamba12,elizalde14,brevik15}, and the recent papers \cite{brevik16,bamba16a} are devoted to inflationary cosmological models with viscous coupled fluids.

In this article we will reproduce inflation by using an inhomogeneous  EoS parameter and formulate the conditions for the existence of inflation without self-reproduction. As it will be shown below, this viscous fluid model is in agreement with the PLANCK satellite data.

\section{Inflationary viscous models without self-reproduction}

We will now investigate which conditions in the very early universe may allow us to avoid the phenomenon of self-reproduction. The conditions are to be expressed in terms of the EoS parameters and of the bulk viscosity. We apply the inhomogeneous viscous fluid formalism to a flat Friedmann-Lemaitre-Robertson-Walker (FLRW) spacetime.

The gravitational field equations acquire the following form,
\begin{align}
& \rho^\prime (N)+3[\rho(N)+P(N)]=0, \notag \\
 &-\frac{2}{k^2}H(N)H^\prime(N)=\rho +P, \label{2}
\end{align}
where $H=\dot{a}/a$ is the Hubble parameter, with $a(t)$ being the scale factor. The dot denotes derivative with respect to the cosmic time $t$; $\rho$ is the density and $P$ is the pressure of the (one-component) fluid. The prime means derivative with respect to the e-folding parameter $N$, i.e. $\rho^\prime (N)=d\rho(N)/dN$ and $H^\prime (N)=dH(N)/dN$.

We now write Friedmann's first equation as
\begin{equation}
\frac{3}{k^2}H^2(N)=\rho, \label{3}
\end{equation}
and drive the  EoS to have the following inhomogeneous form,
\begin{equation}
P(N)=\omega(N)\rho(N)+\zeta(N). \label{4}
\end{equation}
Here $\zeta(N)$ is a {\it function} of the bulk viscosity, and not the viscosity itself (observe that in geometric units the dimension of $\zeta(N)$ is cm$^{-4}$, whereas the dimension of viscosity is cm$^{-3}$). The notation in (\ref{4}) is the same as in the recent paper \cite{brevik16}. The  EoS (\ref{4}) is analogous to the inhomogeneous  EoS often studied in the literature, for instance in \cite{brevik07}, but there without use of the viscosity concept.

We will now study the conditions (\ref{1}) above, for having inflationary models with viscosity and avoiding the self-reproduction issue. One can actually consider many different dependences of the Hubble parameter on $N$ and in fact, with regard to comparison with astronomical data, to determine a specially fitting parametrization of $H$ is a major issue. We feel that the issue we are dealing with is not at this level of accuracy and, therefore, we shall here consider, for the sake of comparison, the three main possible dependences of $H(N)$ on $N$, namely linear, quadratic and exponential.

\subsection{Linear form for the Hubble parameter}

Let us consider the following linear form for $H$ \cite{bamba14},
\begin{equation}
H(N)=G_0N+G_1, \label{5}
\end{equation}
where $G_0<0$ and $G_1>0$ are constant parameters.

We choose the viscosity function $\zeta(N)$ to be proportional to the square of $H$ \cite{brevik16},
\begin{equation}
\zeta(N)=\theta H^2(N), \label{6}
\end{equation}
where $\theta$ is a positive dimensional constant being, in geometric units, its dimension cm$^{-2}$.

Using the gravitational equation of motion (\ref{2}), with (\ref{5}) and (\ref{6}), we obtain the following expression for the thermodynamic parameter
\begin{equation}
\omega(N)=-1-\frac{1}{3}\left[ \frac{2G_0}{H(N)}+k^2\theta \right]. \label{7}
\end{equation}
The second term in the bracket represents the contribution from viscosity. Therefore, the  EoS (\ref{4}) can be written as
\begin{equation}
P(N)=\left[ -1-\frac{2G_0}{H(N)}\right] \rho(N). \label{8}
\end{equation}
We now define the conditions under which the thermodynamic parameter (\ref{7}) in the EoS satisfies all the conditions (\ref{1}) for the avoidance of self-reproduction. From condition (1a), we obtain
\begin{equation}
\theta=\frac{1}{k^2}\left( \frac{2|G_0|}{G_0+G_1}-3\right). \label{9}
\end{equation}
Here $G_1< \frac{5}{3}|G_0|$, because $\theta >0$. From condition (1b), we find the following expression for the e-folding parameter $N_m$ at the beginning of inflation,
\begin{equation}
N_m =\frac{G_1}{|G_0|}-\frac{2}{2+k^2\theta}. \label{10}
\end{equation}
This expression thus determines the relationship between the parameters at the beginning of inflation.

Now consider the condition (1c) for inflation to occur,
\begin{equation}
1<N< \frac{G_1}{|G_0|}-\frac{2}{3+k^2 \theta} \equiv N_m^\prime. \label{11}
\end{equation}
Consequently, the conditions (1b-1c) can be simultaneously satisfied if $1<N<N_m~(N_m < N_m^\prime)$.

Next consider the condition (1d) for "non-self-reproduction" of inflation. Taking into account (\ref{7}), this condition can be simplified, to read
\begin{equation}
9H^3(N)+\theta H(N)+\frac{2G_0}{k^2}<0. \label{12}
\end{equation}
The solution of this inequality is
\begin{equation}
1<N<\frac{1}{|G_0|}\left( G_1-\frac{2}{3}\sqrt{\frac{\theta}{3}} \cot 2\alpha \right), \label{13}
\end{equation}
where
\begin{equation*}
\tan \alpha= \left\{\tan \left[ \frac{1}{2}
\arctan \frac{k^2}{3|G_0|}
\left(\frac{\theta}{3}\right)^{3/2}\right] \right\}^{1/3},      \quad  |\alpha| \leq \frac{\pi}{4}.
\end{equation*}
From comparison of (\ref{10}) and (\ref{13}) it follows that, in order to avoid the initial condition problem, the following, necessary relationship has to be satisfied
\begin{equation}
G_0+G_1=\frac{2}{3}\sqrt{\frac{\theta}{3}}\cot 2\alpha. \label{14}
\end{equation}
Thus, we have obtained for the inflationary model with a linear form for $H$, the form (\ref{8}) for the  EoS. If we want to avoid self-reproduction and the initial condition problem, the thermodynamic parameter (\ref{7}) in the  EoS must simultaneously satisfy the requirements (\ref{9}), (\ref{10}), and (\ref{14}).

\subsection{Exponential form for the Hubble parameter}

In this example we will study the following form for $H$ \cite{bamba14},
\begin{equation}
H(N)=G_2e^{\beta N} +G_3, \label{15}
\end{equation}
where the constant parameters satisfy $G_2<0, G_3>0, \beta>0$. Such an exponential form can describe powerlaw inflation.

Let us suppose that the viscosity function is proportional to $H$ \cite{brevik16},
\begin{equation}
\zeta(N)=\tilde{\theta}H(N), \label{16}
\end{equation}
where $\tilde{\theta}$ is a constant. In geometric units its dimension is cm$^{-3}$. Using the gravitational equation (\ref{2}) for the energy together with (\ref{15}) and (\ref{16}), we find the thermodynamic parameter to have the following form,
\begin{equation}
\omega(N)=-1-\frac{2\beta G_2e^{\beta N}+k^2\tilde{\theta}}{3H(N)}. \label{17}
\end{equation}
The corresponding  EoS becomes
\begin{equation}
P(N)=\left\{ -1-\frac{2\beta [H(N)-G_3]+k^2\tilde{\theta}}{3H(N)}\right\}\rho(N)+\tilde{\theta}H(N). \label{18}
\end{equation}
We will now investigate this inflationary model again with the purpose of avoiding self-reproduction. To start, we use the condition (1a) above to derive, for the parameter $\tilde{\theta}$,
\begin{equation}
\tilde{\theta}=\frac{3}{k^2}\left[ \left( \frac{2}{3}\beta+1\right)e^\beta |G_2|-G_3\right]. \label{19}
\end{equation}
Since $\tilde{\theta}>0$, this means that $G_3 < \left( \frac{2}{3}\beta+1\right)e^\beta |G_2|$. Further, we find from condition (1b) the following e-folding parameter $N$, at the beginning of inflation,
\begin{equation}
N_m^\prime =\frac{1}{\beta}\ln \frac{k^2\tilde{\theta}+2G_3}{2|G_2|(\beta+1)}. \label{20}
\end{equation}
From condition (1c)  the following boundaries for inflation are obtained
\begin{equation}
1<N<\frac{1}{\beta}\ln \left[ \frac{k^2\tilde{\theta}+3G_3}{(2\beta+3)|G_2|}\right] \equiv N_m^{\prime\prime}. \label{21}
\end{equation}
This is necessary in order  to satisfy the conditions (1b-1c), namely $N_m^\prime >N_m^{\prime\prime}$. Inflation without self-reproduction can now be realized if we take into account (\ref{17}) and the condition (1d). Then (1d) simplifies to
\begin{equation}
H^3(N)+\frac{2\beta}{9k^2}H(N)+\frac{1}{9}\left( \tilde{\theta}-\frac{2\beta}{k^2}G_3\right) <0. \label{22}
\end{equation}
The solution (\ref{22}) can be rewritten as
\begin{equation}
1<N<\frac{1}{\beta}\ln \frac{1}{|G_2|}\left[ \left( \frac{2}{3}\right)^{3/2}\frac{\sqrt{\beta}}{k}\cot 2\alpha +G_3\right], \label{23}
\end{equation}
where
\begin{equation*}
\tan \alpha =\left\{\tan \left[ \frac{1}{2}\arctan \frac{1}{\tilde{\theta}-\frac{2\beta}{k^2}G_3}\left( \frac{2}{3}\right)^{5/2}\left( \frac{\sqrt{\beta}}{k}\right)^3 \right] \right\}^{1/3}, \quad |\alpha|\leq \frac{\pi}{4}, \quad \tilde{\theta}<\frac{2\beta}{k^2}G_3.
\end{equation*}
Comparing (\ref{21}) with (\ref{23}) we choose the e-folding parameter $N_m$ at the beginning of inflation to be equal to
\begin{equation}
N_m =\frac{1}{\beta}\ln \left[ \frac{3k^2\tilde{\theta}+G_3}{(2\beta +3)|G_2|}\right], \label{24}
\end{equation}
where
\begin{equation}
G_3=e^\beta -\left(\frac{2}{3}\right)^{3/2}\frac{\sqrt{\beta}}{k}\cot 2\alpha. \label{25}
\end{equation}
We have thus shown that it is possible to have a regime of ``non-self-reproducibility" in the inflationary scenario assuming an exponential behavior for the Hubble parameter. The thermodynamic parameter, in the form (\ref{17}), simultaneously satisfies the conditions (1a-d) provided the equations (\ref{19}), (\ref{24}), and (\ref{25}) hold.

\subsection{Quadratic form for the Hubble parameter}
Here we  consider a differet dependence for $H$, \cite{bamba14}
\begin{equation}
H(N)=G_4N^2+G_5, \label{26}
\end{equation}
where $G_4<0$ and $G_5>0$ are constant parameters. The viscosity function $\zeta(H)$  has, again, the form (\ref{6}). From the gravitational equation (\ref{2}), together with (\ref{6}) and (\ref{26}), we obtain for the thermodynamic parameter
\begin{equation}
\omega(N)=-1-\frac{1}{3}\left[ \frac{4G_4N}{H(N)}+k^2\theta\right]. \label{27}
\end{equation}
The consequence being that the thermodynamic parameter contains a contribution from viscosity.

The corresponding  EoS becomes
\begin{equation}
P(N)=\left\{ -1-\frac{4}{3}\left[ \frac{G_4N}{H(N)}+\frac{1}{3}k^2\theta \right]\right\} \rho(N). \label{28}
\end{equation}
Analogously to our previous considerations, we will check how the present inflationary model fits with the conditions (1a-d). From condition (1a) we find the following expression for the parameter $\theta$,
\begin{equation}
\theta=\frac{4|G_4|}{k^2(G_4+G_5)}, \label{29}
\end{equation}
while condition (1b) yields the e-folding parameter $N$, at the beginning of inflation, to be
\begin{equation}
N_m=-\frac{2}{2+k^2\theta}+\sqrt{ \frac{4}{(2+k^2\theta)^2}+\frac{G_5}{|G_4|}}. \label{30}
\end{equation}
Condition (1c) results in
\begin{equation}
1<N<-2(3+k^2\theta)+\sqrt{ 4(3+k^2\theta)^2+\frac{G_5}{|G_4|}} = N_m^\prime. \label{31}
\end{equation}
In order to fulfill the conditions (1b-c) simultaneously, we must require that $1<N<N_m$ $(N_m<N_m^\prime)$. The last condition (1d), together with (\ref{7}), leads to the inequality
\begin{equation}
H^3(N)+\frac{\theta}{9}H(N)+\frac{4G_4}{9k^2}N<0. \label{32}
\end{equation}
It has the solution
\begin{equation}
1<N<\frac{1}{|G_4|}\sqrt{ \frac{2}{3}\sqrt{\frac{\theta}{3}}\,|\cot 2\alpha| +G_5}, \label{33}
\end{equation}
where
\begin{equation*}
\tan \alpha=\left\{ \tan \left[ \frac{1}{2}\arctan \frac{k^2}{18 |G_4|}\sqrt{\frac{\theta^3}{3}}\right] \right\}^{1/3}, \quad |\alpha| \leq \frac{\pi}{4}.
\end{equation*}
From a comparison of (\ref{30}) and (\ref{33}) we see that, in order to avoid the initial condition problem, it is necessary that the following relationship between the parameters holds
\begin{equation}
\gamma+\sqrt{\gamma^2+\frac{G_5}{|G_4|}} = \frac{1}{3\gamma}\sqrt{\frac{\theta}{3}}\, \Big| \frac{\cot 2\alpha}{G_4}\Big|, \label{34}
\end{equation}
with $\gamma=\frac{2}{2+k^2\theta}$.

Thus, these inflationary models do realize the existence of the regime of ``non-self-reproduction", provided the relations (\ref{29}), (\ref{33}), and (\ref{34}) hold, for the thermodynamic parameter of the form (\ref{27}).

\section{Comparison of our inflationary models with observational data}

In this section we will calculate the inflationary parameters and consider, in particular,  the matching of the spectral index and tensor-to-scalar ratio with the most recent data available from the PLANCK surveyor.

Let us first calculate the ``slow-roll" slope parameter \cite{myrzakulov15}
\begin{equation}
\varepsilon = -\frac{\dot{H}}{H^2}. \label{35}
\end{equation}
Assuming the linear form (\ref{5}) for the Hubble parameter, one gets
\begin{equation}
\varepsilon =\frac{|G_0|}{G_0N+G_1}. \label{36}
\end{equation}
In order to have acceleration, one must require $\varepsilon <1$.  In our case, this corresponds to
\begin{equation}
1<N<\frac{G_1-|G_0|}{|G_0|}. \label{37}
\end{equation}
Another important ``slow-roll" parameter is \cite{myrzakulov15}
\begin{equation}
\eta=\varepsilon-\frac{1}{2\varepsilon H} \dot{\varepsilon}. \label{38}
\end{equation}
In our case, $\eta=\varepsilon/2$, and the power spectrum is \cite{myrzakulov15}
\begin{equation}
\Delta_R^2=\frac{k^2H^2}{8\pi^2\varepsilon}. \label{39}
\end{equation}
From Eqs. (\ref{5}) and (\ref{36}), we find
\begin{equation}
\Delta_R^2=\frac{k^2(G_0N+G_1)^3}{8\pi |G_0|}. \label{40}
\end{equation}
And taking into account the slow-roll parameters, one can calculate the spectral index $n_s$ and the tensor-to-scalar ratio $r$, as
\begin{equation}
n_s=1-6\varepsilon+2\eta, \quad  r=16\varepsilon. \label{41}
\end{equation}
We then obtain
\begin{equation}
n_s=1-\frac{5|G_0|}{G_0N+G_1}, \quad r=\frac{16|G_0|}{G_0N+G_1}. \label{42}
\end{equation}
From the PLANCK satellite 2015 astronomical results, we know that $n_s=0.9603 \pm 0.0073$. To be in accordance with this result, we require that $\frac{|G_0|}{G_0N+G_1}=0.00794 \pm 0.000146.$

The relation between the EoS parameter and the tensor-to-scalar ratio is
\begin{equation}
\omega(N)=-1+\frac{r}{24}-\frac{1}{3}k^2\theta, \label{43}
\end{equation}
where the influence from viscosity is present in the last term. A similar relation for the inflationary universe for a perfect fluid without viscosity was explored in \cite{bamba14}.

On the other hand, assuming the exponential form (\ref{15}) for the Hubble parameter, we obtain
\begin{equation}
\varepsilon = \beta |G_2|\frac{e^{\beta N}}{G_2e^{\beta N}+G_3}. \label{44}
\end{equation}
The regime of acceleration corresponds to
\begin{equation}
1<N<\frac{1}{\beta}\ln \frac{G_3}{(\beta+1)|G_2|}, \label{45}
\end{equation}
and one  calculates from here
\begin{equation}
\eta=\varepsilon \left( 1-\frac{G_3}{|G_2|}e^{-\beta N}\right). \label{46}
\end{equation}
The amplitude of the primordial scalar power spectrum (\ref{39}) is given by
\begin{equation}
\Delta_R^2=\frac{k^2(G_2e^{\beta N}+G_3)^3}{8\pi^2\beta |G_2|e^{\beta N}}. \label{47}
\end{equation}
From the slow-roll parameters one gets for the spectral index, $n_s$, and the tensor-to-scalar ratio, $r$,
\begin{equation}
n_s=1+2\varepsilon \left( \frac{G_3}{G_2}e^{-\beta N}-2\right), \quad r=16\beta |G_2|\frac{e^{\beta N}}{G_2e^{\beta N}+G_3}. \label{48}
\end{equation}
Thus, the relationship between the thermodynamic parameter $\omega$ and the tensor-to-scalar ratio $r$ becomes
\begin{equation}
\omega(N)=-1+\frac{r}{24}-\frac{1}{3}\frac{k^2\tilde{\theta}}{H(N)}. \label{49}
\end{equation}
Finally, for the square of the Hubble parameter (\ref{26}) we have
\begin{equation}
\varepsilon =\frac{2|G_4|N}{G_4N^2+G_5}, \quad \eta=\frac{3}{2}\varepsilon +\frac{1}{2N}. \label{50}
\end{equation}
The acceleration corresponds to
\begin{equation}
1<N<-1+\sqrt{1+\frac{G_5}{|G_4|}}, \label{51}
\end{equation}
and the power spectrum is given by
\begin{equation}
\Delta_R^2=\frac{k^2(G_4N^2+G_5)^3}{16\pi |G_4|N}. \label{52}
\end{equation}
We can write, for our designed observables in this inflationary model,
\begin{equation}
n_s=1-3\varepsilon +\frac{1}{N}, \quad r=\frac{32|G_4|N}{G_4N^2+G_5}. \label{53}
\end{equation}
Finally, the thermodynamic parameter $\omega$ can be expressed in terms of $r$ and $\theta$,
similarly as in (\ref{43}).

\section{Conclusions}

We have considered in this paper inflation from fluid models which take into account bulk viscosity, in a FLRW spacetime. We have searched for the conditions  that tell us how to avoid the self-reproduction issue in the very early universe, namely at about $10^{-33}~$s after the big bang. We have studied in detail three different inflationary models that allow us to do that. Specifically, in terms of the e-folding parameter $N$ we  have discussed the cases where the Hubble parameter $H(N)$ varies: (i) linearly with $N$, (ii) exponentially as $e^{\beta N}$ with $\beta$ a positive constant, and (iii) quadratically with $N$. Expressions for the thermodynamic parameter $\omega(N)$ in the  EoS in terms of $N$ have been obtained for all these models.

Moreover, for all those models, the corresponding conditions leadingto ``non-self-reproduction" have been analyzed, following the method outlined in \cite{mukhanov14}. It has been shown that these fluid models are  compatible with an inflationary universe. Expressions for the spectral index $n_s$, the tensor-to-scalar ratio $r$, and the power spectrum have been explicitly calculated. Also, we have derived useful expressions for the thermodynamic parameter in terms of the tensor-to-scalar ratio and the viscosity function. Agreement with the most recent and accurate astronomical data, obtained with the PLANCK satellite surveyor, has been demonstrated.

\bigskip

\noindent{\bf Acknowledgments}.
 This work has been supported by  grants from the Russian Ministry of Education and Science, Project TSPU-139 (A.V.T.), and from MINECO (Spain), Project FIS2013-44881, and CSIC, project I-LINK 1019 (E.E.).
\bigskip

 \noindent{\large \bf Appendix}.
  \medskip

Let us consider here explicitly the following generalization of the linear form for the Hubble parameter:
\begin{equation}
H(N)=G_0N^n+G_1, \label{app1}
\end{equation}
where the parameter $n$ is a positive integer, while $G_0 <0$ and  $G_1 >0$.

We will check the possible agreement of this theoretical model for general $n$ with the PLANCK
satellite observational data. To start, we calculate the corresponding slow-roll parameters:
\begin{equation}
\varepsilon =\frac{n |G_0| N^{n-1}}{H}, \qquad \eta=\frac{1}{2}\left( \varepsilon -\frac{n-1}{N}\right). \label{app2}
\end{equation}
The regime of universe acceleration will hold whenever this inequality is fulfilled: $\varepsilon <1$. In our case this  is equivalent to the following condition:
\begin{equation}
N^n+nN^{n-1} - \frac{G_1}{|G_0|} < 0. \label{app3}
\end{equation}
In the particular case when $n=2$,  we obtain the solution
\begin{equation}
1<N< -1+ \sqrt{1+\frac{G_1}{|G_0|}}. \label{app4}
\end{equation}
If we consider the asymptotic case, where $N \to 1$. Then, in (\ref{app3}) the second
term yields the main contribution, and the inequality (\ref{app3}) simplifies as
\begin{equation}
nN^{n-1}< \frac{G_1}{|G_0|}. \label{app5}
\end{equation}
Hence, it follows that $ N <[G_1/(n|G_0|)]^{1/(n-1)}$, $n>1$, and
the e-folding parameter $N$ changes in the region $1\approx N <[G_1/(n|G_0|)]^{1/(n-1)}$, $n>1$.
On the contrary, when $N$ is larger (recall that about 50-70 e-foldings are required to solve
the horizon and flatness problems), then in (\ref{app3}) it is the first
term that yields the main contribution, and the inequality (\ref{app3}) reads now
\begin{equation}
N^n< \frac{G_1}{|G_0|}; \label{app5b}
\end{equation}
the e-folding parameter $N$ changes in the region $1 << N <(G_1/|G_0|)^{1/n}$, in this case.

Finally, we find the power spectrum:
\begin{equation}
\Delta_R^2=\frac{k^2(G_0N^n +G_1)^3}{8\pi n |G_0| N^{n-1}}. \label{app6}
\end{equation}
The spectral index $n_s$ and the tensor-to-scalar ratio $r$ turn out to be
\begin{equation}
n_s=1-5\varepsilon - \frac{n-1}{N}, \quad r=\frac{16n |G_0| N^{n-1}}{G_0N^n+G_1}. \label{app7}
\end{equation}
Further, we now discuss the coincidence of this inflationary model with PLANCK's astronomical results. In order to reproduce the most recent observations, it is necessary to demand that
\begin{equation}
\frac{(4n+1) |G_0| N^n + (n-1)G_1}{N(G_0N^n+G_1)} = 0.0397\pm 0.0073,  \label{app8}
\end{equation}
and another restriction we encounter reads
 \begin{equation}
\frac{n |G_0| N^{n-1}}{G_0N^n+G_1} < 0.006875.  \label{app9}
\end{equation}
All these restrictions can be easily accounted for and
thus, we conclude that our model is able to describe the evolution of the inflationary universe.


\begin{thebibliography}{99}


\bibitem{ade13} P. A. R. Ade {\it et al.} [Planck Collaboration], arXiv: 1303.5076 [astro-ph.CO].

\bibitem{ade13a}  P. A. R. Ade {\it et al.} [Planck Collaboration], arXiv: 1303.5076 [astro-ph.CO].

\bibitem{mukhanov14} V. Mukhanov, Fortschritte der Physik {\bf 63}, 1 (2014), arXiv: 1409.2335 [astro-ph.CO].

\bibitem{nojiri15} S. Nojiri and  S. D. Odintsov, Astrophys. Space Sci. {\bf 357}, 39 (2015), arXiv: 1412.2518 [gr-qc].

\bibitem{bamba:2015uma} K. Bamba and S. D. Odintsov, Symmetry {\bf 7}, No. 1, 220 (2015), arXiv: 1503.00442 [hep-th].



\bibitem{bamba14} K. Bamba, S. Nojiri, S. D. Odintsov and D. Saez-Gomez, Phys. Rev. D {\bf 90}, 124061 (2014), arXiv: 1410.3993 [hep-th].

\bibitem{linde} A. Linde, arXiv: 1402.0526 [hep-th].

\bibitem{addrefs1} M. M. Disconzi, K. W. Thomas and R. J. Scherrer,  Phys. Rev. D {\bf 91}, 04532 (2015);
I. Brevik, E. Elizalde, S. Nojiri, and S. D. Odintsov. Phys. Rev. D {\bf 84}, 103508 (2011);
B. Giacomazzo and L. Rezzolla,  J. Fluid Mech. {\bf 562}, 223 (2006);
W. Zimdahl,  Mon. Not. R. Astron. Soc. {\bf 280}, 1239 (1996);
J. D. Barrow,  Phys. Lett. B {\bf 180}, 335 (1986);  A. Lichnerowicz,  J. Math. Phys. {\bf 17}, 2135 (1976).

\bibitem{brevik06} I. Brevik, J. M. B{\o}rven and S. Ng, Gen. Relativ. Grav. {\bf 38}, 907 (2006).

\bibitem{nojiri05} S. Nojiri and S. D. Odintsov, Phys. Rev. D {\bf 72}, 023003 (2005).

\bibitem{brevik10} I. Brevik, S. Nojiri, S. D. Odintsov and D. Saez-Gomez, Eur. Phys. J. C {\bf 69}, 563 (2010).

\bibitem{brevik14} I. Brevik, V. V. Obukhov and A. V. Timoshkin, Astrophys. Space Sci. {\bf 355}, 2163 (2014).


\bibitem{myrzakulov15} R. Myrzakulov and L. Sebastiani, Astrophys. Space Sci. {\bf 356}, 205  (2015), arXiv: 1410.3573v2 [gr-qc].

\bibitem{capozziello06} S. Capozziello, V. F. Cardone, E. Elizalde, S. Nojiri and S. D. Odintsov, Phys. Rev. {\bf 73}, 043512 (2006).

\bibitem{nojiri06} S. Nojiri and S. D. Odintsov, Phys. Lett. B {\bf 639}, 144 (2006), arXiv: hep-th/0606025.

\bibitem{nojiri07} S. Nojiri and S. D. Odintsov, Phys. Lett. B {\bf 649}, 440 (2007), arXiv: hep-th/0702031.





\bibitem{nojiri07a} S. Nojiri and S. D. Odintsov, J. Geom. Meth. Mod. Phys. {\bf 4}, 115 (2007), arXiv: hep-th/0601213.

\bibitem{nojiri11} S. Nojiri and S. D. Odintsov, Phys. Rep. {\bf 505}, 59 (2011), arXiv: 1011.0544.

\bibitem{bamba12} K. Bamba, S. Capozziello, S. Nojiri and S. D. Odintsov, Astrophys. Space Sci. {\bf 342}, 155 (2012), arXiv: 1205.3421 [gr-qc].

\bibitem{elizalde14} E. Elizalde, V. V. Obukhov and A. V. Timoshkin, Mod. Phys. Lett. A {\bf 29}, 1450132 (2014).

\bibitem{brevik15} I. Brevik, V. V. Obukhov and A. V. Timoshkin, Astrophys. Space Sci. {\bf 355}, 399 (2015).

\bibitem{brevik16} I. Brevik and A. V. Timoshkin, J. Exp. Theor. Phys. (JETP) {\bf 122}, 679 (2016).

\bibitem{bamba16a} K. Bamba and S. D. Odintsov, Eur. Phys. J C {\bf 76}, No. 1, 18 (2016), arXiv: 1508.05451 [gr-qc].

\bibitem{brevik07} I. Brevik, E. Elizalde, O. Gorbunova and A. V. Timoshkin, Eur. Phys. J. C {\bf 52}, 223 (2007).

\end{thebibliography}
\end{document}